\documentclass[12pt]{article}
\usepackage{amsmath,amssymb,bm}
\usepackage{cite}   
\usepackage{hyperref}
\hypersetup{colorlinks=true}
\title{Hawking radiation via tunneling from the spacetime of a spinning cosmic string black holes }
\date{}
\author{\bf Kimet Jusufi\footnote{E-mail: kimet.jusufi@unite.edu.mk}}

\makeatletter
 
  \@addtoreset{equation}{section}
 \makeatother

\begin{document}
\maketitle

\centerline{\it Department of Physics, State University of Tetovo, Ilinden Street nn, 1200, Macedonia}

\vskip 0.5 truecm

\begin{abstract}
In this paper, we study Hawking radiation as a massless particles tunneling
process across the event horizon from the Schwarzschild and Reissner-Nordstr\"{o}m black holes pierced by an infinitely long spinning cosmic string and
a global monopole. Applying the WKB approximation and using a generalized
Painlev\'{e} line element for stationary axisymmetric spacetimes, also by taking into
account that the ADM mass of the black hole decreases due to the presence of
topological defects, it is shown that the Hawking temperature remains unchanged
for these black holes. The tunneling of charged massive particles from Reissner-Nordstr\"{o}m black holes is also studied, in both cases the tunneling rate is related to the change of the Bekenstein-Hawking entropy. The results extend the work of Parikh and
Wilczek and are consistent with an underlying unitary theory.

%\PACS{PACS code1 \and PACS code2 \and more}
\end{abstract}

\section{Introduction}

Combining the principles of quantum field and general relativity leads to the emission of thermal radiation from a black hole, known as Hawking radiation. This amazing discovery by Steven Hawking \cite{hawking}, suggest that black holes could lose energy, shrink, and as a consequence evaporate completely. Stated in other words, one can start with some pure quantum state falling into a black hole and end up after the evaporation with a thermal state which is a mixed state. This implies the loss of information, also known as the ``information loss paradox'' which continues to be unsolved.

Although many derivations of this radiation have been made \cite{hawking,gibbons,perkih,umetsu}, recently Perkih and Wilczek treated Hawking radiation as a tunneling process using a semi classical WKB approximation \cite{perkih}. According to this picture  a particle-antiparticle pair is formed close to the horizon, the particle with positive energy  can quantum mechanically tunnel through the horizon and it is observed at infinity as the Hawking flux. A number of spherically symmetric and stationary black holes have been studied, for scalar as well as for Dirac particles. In particular, the Kerr black hole, Kerr-Newman black hole \cite{zhang,jiang}, black hole in spacetime with topological defects \cite{ren,vilenkin}. The aim of this paper, is to extend the results presented here \cite{ren}, and investigate the tunneling process and the Hawking temperature for the Schwarzschild and Reissner-Nordstr\"{o}m black holes spacetimes with an infinitely long spinning cosmic string and a global monopole. In order to calculate the Hawking temperature, we need to remove the coordinate singularity at the horizon. Fortunately, there exists a generalization of the Painlev\'{e} coordinates to stationary axisymmetric space-times \cite{zhang}, which will be used in this paper. 

Cosmic strings are one dimensional object, that may have been produced by the phase transition in the early universe. The spinning cosmic string is characterized by the rational parameter $a$ and the angular parameter $J$, given by $a =4J$. The space-time of a cosmic string has some quite surprising features; the spacetime is locally flat, but globally is conical. One can think about this as locally flat space-time with a wedge removed given by the deficit angle $\delta=8\,\pi \,G \,\mu $. On the other hand, the spacetime geometry of a global monopole of the surface $\theta=\pi/2$, is conical, similar to a cosmic string  with deficit angle $\delta \Omega=8\pi^{2}G\eta^{2}$.

The structure of this article is therefore as follows. In Section 2, we briefly review and introduce analogue Painlev\'{e}-Schwarzschild line element in spacetime with a spinning cosmic string and a global monopole. In Section 3, we calculate the tunneling rate of massless particles and derive the Hawking temperature for Schwarzschild black hole. In Section 4, we derive the tunneling rate and Hawking temperature for Reissner-Nordstr\"{o}m cosmic string black holes. In section 5, we derive the tunneling rate of charged massive particles for Reissner-Nordstr\"{o}m cosmic string black holes. In Section 6, we comment on our results. We assume natural units in which $c=\hbar=k=G=1$ and metric signature $(-,+,+,+)$ throughout the paper.

\section{Painlev\'{e} coordinates for a cosmic string black hole}
Let us begin by writing the line element of a black hole pierced by an infinitely long cosmic string and a global monopole in spherical coordinates\cite{vilenkin1}
\begin{equation}
\mathrm{d}s^{2}=-\chi^{2}(\mathrm{d}t_{s} +a \mathrm{d}\varphi)^2+\chi^{-2}\mathrm{d}r^{2}+r^2p^{2}(\mathrm{d}\theta^2+b^2\sin^2\theta \mathrm{d}\varphi^2)
\label{metric1}
\end{equation}
where $\chi=\sqrt{1-2M/r}$ is  the redshift factor, $b=(1-4\,\mu)$ the cosmic string parameter which belongs to the interval $(0,1]$, $p^{2}=1-8\pi\eta^{2}$ encodes the presence of a global monopole and $a=4\,J$ is the rational parameter of cosmic string. The metric \eqref{metric1} has a coordinate singularity at $r_{H}=2M$, which can be removed. Starting from the general form of the spacetime metric \eqref{metric1} and taking into account the dragging angular velocity $\Omega_{b,p}=\dot{\varphi}=-g_{03}/g_{33}$, than the stationary spacetime metric \eqref{metric1} can be rewritten as
\begin{align}
\mathrm{d}s^{2}&=-\frac{r^{2}b^{2}p^{2}\sin^{2}\theta \,\chi^{2}}{r^{2}b^{2}p^{2}\sin^{2}\theta-a^{2}\chi^{2}}\mathrm{d}t_{s}^{2}+\chi^{-2}\mathrm{d}r^{2}+r^{2}\mathrm{d}\theta^{2}
\end{align}
This metric represents a three-dimensional hypersurface in the four-dimensional cosmic string black hole spacetime. Performing a further appropriate coordinate transformation, such that the event horizon to coincide with the infinite red-shift surface, given by
\begin{equation}
\mathrm{d}t_{s}=\mathrm{d}t+F(r,\theta)\mathrm{d}r+G(r,\theta)\mathrm{d}\theta,
\end{equation}
where $F(r,\theta)$ and $G(r,\theta)$ are two functions to be determined and satisfying the integrability conditions $ \partial F(r,\theta)/\partial \theta=\partial G(r,\theta)/\partial r$. Next, by demanding that constant-time slices are flat Euclidean space in radial by impossing the following condition $g_{11}+\hat{g}_{00}F^{2}(r,\theta)=1$, where $\hat{g}_{00}=g_{00}-(g_{03})^{2}/g_{33}$, than one can find a general axisymmetric Painliev\'{e} line element \cite{Zhang1}
\begin{eqnarray}\nonumber
\mathrm{d}s^{2}
& = & \hat{g}_{00}\mathrm{d}t^{2}+2\sqrt{\hat{g}_{00}(1-g_{11})}\mathrm{d}t\,\mathrm{d}r +\mathrm{d}r^{2}\\\nonumber
& + &\left[\hat{g}_{00}G(r,\theta)^{2}+g_{22}\right]\mathrm{d}\theta^{2}+2\hat{g}_{00}G(r,\theta)\mathrm{d}t\,\mathrm{d}\theta\\
& + & 2\sqrt{\hat{g}_{00}(1-g_{11})}G(r,\theta)\mathrm{d}t\,\mathrm{d}\,\theta.
\label{metric2}
\end{eqnarray}
This metric has several attractive properties: It is well behaved at the event horizon $r_{H}=2M$, the coordinate $t$ represent the local proper time for radially free falling obsorver, constant-time slices are just flat Euclidian space in radial, and finally this metric satisfies Landau's condition of coordinate clock synchronization (see, e.g., \cite{zhang}). Using the metric \eqref{metric1} and \eqref{metric2}, it follows that the Painlev\'{e}-Schwarzschild cosmic string black hole line element for constant $\theta=\pi/2$ is given by
\begin{align}\nonumber
\mathrm{d}s^{2}&=-\frac{r^{2}b^{2}p^{2}\,\chi^{2}}{r^{2}b^{2}p^{2}-a^{2}\chi^{2}}\mathrm{d}t^{2}+2\sqrt{\frac{2Mrb^{2}p^{2}}{r^{2}b^{2}p^{2}-a^{2}\chi^{2}}}\,\mathrm{d}t\,\mathrm{d}r+\mathrm{d}r^{2}.
\label{m1}
\end{align}

Solving for the radial null geodesics $ \mathrm{d}\theta=\mathrm{d}s^{2}=0$ one can show that
\begin{equation}
\dot{r}\equiv\frac{\mathrm{d}r}{\mathrm{d}t}=\frac{rbp}{\sqrt{r^{2}b^{2}p^{2} -a^{2}(1-\frac{2M}{r})}}\left(\pm 1-\sqrt{\frac{2M}{r}}\right),
\label{geodesic}
\end{equation}
where the $+(-)$ sign correspondts to an outgoning (ingoing) geodesics, respectively. In the particular case, $b=1, p=1$ and $a=0$, Eq.\eqref{geodesic} reduces to Painlev\'{e}- Schwarzschild line element without topological defects. 

\section{Particles tunneling from Schwarzschild black hole}
The tunneling rate is related to the imaginary part of the action in the classically forbidden region
\begin{equation}
\Gamma \sim e^{-2\,\text{Im}\, S}.
\label{tunneling}
\end{equation}

Since the metric \eqref{metric2} is independent of the coordinate $\varphi $, we can completely eliminate this degree of freedom by writing the action in the form
\begin{equation}
S=\int_{t_{i}}^{t_{f}}(L-P_{\varphi}\dot{\varphi})\,\mathrm{d}t,
\end{equation}
\label{act}
by taking the imaginary part of the action, it follows
\begin{align}\nonumber
\mbox{Im}\,S&=\text{Im}\int_{r_{in}}^{r_{out}}\,\left(P_{r}\dot{r}-P_{\varphi}\dot{\varphi}\right)\frac{\mathrm{d}r}{\dot{r}}\\
&=\text{Im}\int_{r_{in}}^{r_{out}}\left(\int_{(0,0)}^{(P_{r},P_{\varphi})}\,\dot{r}\,\mathrm{d}P'_{r}-\dot{\varphi}\,dP'_{\varphi}\right)\,\frac{\mathrm{d}r}{\dot{r}}.
\label{action}
\end{align}

These equations are modified when the particle’s self-gravitation is taken into account\cite{kraus}. Fixing the ADM mass of the total background spacetime and allowing them to fluctuate, when a particle with a shell of energy $\omega$ is tunneld out, the mass of the black hole and angular momentum will be reduced to $M'\to M-\omega$ and $J'\to (M-\omega)a$, respectively. On the other hand, due to the presence of a cosmic string the ADM mass $M_{b,p}$, energy $E_{b,p}$, and angular momentum $J_{b,p}$ of black hole are decreased by a factor of $(b\,p^{2})$ \cite{ren,vilenkin}
\begin{equation}
M_{b,p}=p^{2}\,b\,M, \,\,E_{b,p}=p^{2}\,b\,E,\,\,J_{b,p}=p^{2}\,b\,J.
\label{mass}
\end{equation} 

Of course the energy of the shell $\omega$ should be also replaced by $p^{2}\,b\,\omega$. So, when the particle tunnels out, the ADM mass becomes $bp^{2}(M-\omega)$, and the angular momentum $bp^{2}a(M-\omega)$, respectively. We can now use the Hamilton's equation
\begin{align}
\dot{r}&=\frac{\mathrm{d}H}{\mathrm{d}P_{r}}\bigg|_{(r;\varphi,P_{\varphi})}=bp^{2}\,\frac{\mathrm{d}(M-\omega)}{\mathrm{d}P_{r}}=bp^{2}\,\frac{\mathrm{d}M'}{\mathrm{d}P_{r}},\\
\dot{\varphi}&=\frac{\mathrm{d}H}{\mathrm{d}P_{\varphi}}\bigg|_{(\varphi;r,P_{r})}=bp^{2}\,a\Omega'_{H}\frac{\mathrm{d}(M-\omega)}{\mathrm{d}P_{\varphi}}=bp^{2}\,a\Omega'_{H}\frac{\mathrm{d}M'}{\mathrm{d}P_{\varphi}},
\end{align}
where $\mathrm{d}H_{\varphi;r,P_{r}}=\Omega'_{H}\mathrm{d}J=bp^{2}\,a\Omega'_{H}\mathrm{d}(M-\omega)$. On the other hand, the dragging angular velocity of the particle at the horizon vanishes
\begin{equation}
\Omega'_{H,bp}=\frac{a\,\tilde{\chi}^{2}(r_{H})}{r^{2}b^{2}p^{2}-a^{2}\,\tilde{\chi}^{2}(r_{H})}\bigg|_{r_{H}=2(M-\omega)}=0.
\label{dragg}
\end{equation}
where $\tilde{\chi}^{2}=1-2(M-\omega)/r$. Therefore, we only have to work out the radial part of the action given by
\begin{align}
\text{Im}\,S&=\text{Im}\int_{bp^{2}M}^{bp^{2}(M-\omega)}\int_{r_{in}}^{r_{out}}\,\frac{\mathrm{d}r}{\dot{r}}\,\mathrm{d}H.
\label{actio1}
\end{align}

Now, since the particles tunnel out, the energy will be $\mathrm{d}H=b\,p^{2}\,\mathrm{d}(M-\omega)=b\,p^{2}\,\mathrm{d}M'$, which implies that we must choose the positive sign in Eq.\eqref{geodesic}, than the action reads
\begin{align}
\mbox{Im}\,S=\mbox{Im}\,\int_{bp^{2}M}^{bp^{2}(M-\omega)}\int_{r_{in}}^{r_{out}}\frac{\sqrt{r^{2}b^{2}p^{2}-a^{2}\,\tilde{\chi}^{2}(r)}}{bp\left(r-\sqrt{2M'r}\right)}\mathrm{d}(bp^{2}M')\mathrm{d}r,
\end{align}
here we regard $\omega$ as a constant and $M'$ as a variable. Using the Feynman's prescription $i \epsilon$ for positive energy solution $\omega \to \omega-i\epsilon$, we obtain $M\to M-i\epsilon$, so that the  last equation can be rewritten as
\begin{align}
\text{Im}\,S&=\text{Im}\int_{r_{in}}^{r_{out}}\int_{bp^{2}\,M}^{bp^{2}\,(M-\omega)}\frac{1}{1-f(M')+i\epsilon}\,\mathrm{d}(bp^{2}M')\mathrm{d}r,
\label{Ims}
\end{align}
and the functions $f(M')$ is given by
\begin{equation}
f(M')=\frac{\sqrt{r^{2}b^{2}p^{2}-a^{2}\tilde{\chi}^{2}\left(r\right)}-bp\left(r-\sqrt{2M'r}\right)}{\sqrt{r^{2}b^{2}p^{2}-a^{2}\tilde{\chi}^{2}(r)}}.
\end{equation}

Here, the mass ranges from $bp^{2}M$ to $bp^{2}(M-\omega)$, wile the radial coordinates ranges from $r_{in}=2M$ to $r_{out}=2(M-\omega)$. Than  Eq.\eqref{Ims} can be written as
\begin{align}
\text{Im}\,S &=\text{Im}\int_{r_{in}}^{r_{out}}\left[\mathcal{P}\frac{1}{1-f(M')}+\int_{bp^{2}M}^{bp^{2}(M-\omega)}\,-i\pi \delta \left(1-f(M')\right)\,\mathrm{d}(bp^{2}M')\right]\mathrm{d}r
\end{align}
where $\mathcal{P}$ stands for the principal value. Since we are interested only for the imagionary part, we can ignore the first term because it is real, and find
\begin{align}\nonumber
\text{Im}\,S&=\text{Im}\left[\int_{r_{in}}^{r_{out}}\left(\int_{bp^{2}M}^{bp^{2}(M-\omega)}\,-i\pi \delta \left(1-f(M')\right)\,\mathrm{d}(bp^{2}M')\right)\mathrm{d}r\right]\\\nonumber
&=bp^{2}\int_{2\,M}^{2\,(M-\omega)}\left(-\pi\,r\right)\mathrm{d}r\\
&=4\pi b p^{2}\omega \left(M-\frac{\omega}{2}\right).
\end{align}

Substituting this result into the Eq.\eqref{tunneling}, the tunneling probability of outgoing particles in terms of the change of the entropy of the black hole will be
\begin{equation}
\Gamma \sim e^{-2\,\mbox{Im}\,S}=\exp{\left[-8\pi bp^{2} M\omega\left(1-\frac{\omega}{2M}\right)\right]}=\exp\left(\Delta S_{BH,bp}\right),
\label{19}
\end{equation}
where $\Delta S_{BH,bp}=S_{BH,bp}(M-\omega)-S_{BH,bp}(M)$, is the change of Bekenstein-Hawking entropy before and after the emission. Although the entropy and surface area depend on the parameters $bp^{2}$, their relation remains unchanged. Neglecting the quadratic term in Eq. \eqref{19}, since $ M>>\omega$, it follows
\begin{equation}
\Gamma\sim \exp\left(-8\pi \,b\,p^{2}\,\omega M\right).
\end{equation}

The energy of the particles in this spacetime also decreases by a factor of $bp^{2}$, and therefore the Boltzmann factor will be $ \Gamma\sim \exp\left(-\beta_{H} bp^{2} \omega\right)$, where $\beta_{H}=1/T$. The Hawking temperature will be
\begin{equation}
T_{H}=\frac{1}{8\pi M}. \label{sh1}
\end{equation}
As a result, we have shown that the Hawking temperature for Schwarzschild black hole in spacetime with a spinning cosmic string and a global monopole  remains unchanged as seen at infinity. 

\section{Tunneling from Reissner-Nordstr\"{o}m  black hole}
Recall that line element for the Reissner-Nordstr\"{o}m black hole pierced by an infintly long cosmic string and a global monopole reads
\begin{equation}
\mathrm{d}s^{2}=-\Delta(\mathrm{d}t +a \,\mathrm{d}\varphi)^2+\Delta^{-1}\mathrm{d}r^{2}+r^2p^{2}(\mathrm{d}\theta^2+b^2\sin^2\theta \mathrm{d}\varphi^2)
\label{Reissner}
\end{equation}
where the function $\Delta(r)$ is given by
\begin{equation}
\Delta\left(r\right)=1-\frac{2M}{r}+\frac{Q^{2}}{r^{2}}=\frac{(r-r_{+})(r-r_{-})}{r^{2}}
\end{equation}
where $r_{\pm}=M\pm\sqrt{M^{2}-Q^{2}}$ gives the distance from the center of the black hole to the horizon. Using  Eqs. \eqref{metric2} and \eqref{Reissner} one can easy find the corresponding Painlev\'{e} type coordinates, simply by replacing $2Mr$ with $2Mr-Q^{2}$, yielding
\begin{eqnarray}
\mathrm{d}s^{2}= -\frac{r^{2}b^{2}p^{2}\Delta(r)}{r^{2}b^{2}p^{2}-a^{2}\Delta(r)}\mathrm{d}t^{2}+2\sqrt{\frac{ b^{2}p^{2}\left(2Mr-Q^{2}\right)}{r^{2}b^{2}p^{2}-a^{2}\Delta(r)}}\,\mathrm{d}t\,\mathrm{d}r+\mathrm{d}r^{2}.
\label{r-n}
\end{eqnarray}

This metric is well behaved at  $r_{+}=M+\sqrt{M^{2}-Q^{2}}$. Solving for the radial null geodesics for an outgoing massless particle  $ \mathrm{d}\theta=\mathrm{d}s^{2}=0$, yields
\begin{equation}
\dot{r}=\frac{rbp}{\sqrt{r^{2}b^{2}p^{2}-a^{2}\Delta(r)}}\left(\pm 1-\sqrt{\frac{2M}{r}-\frac{Q^{2}}{r^{2}}}\right),
\label{g2}
\end{equation}
setting $b=1, a=0$ and $p=1$ into the \eqref{g2} we recover the Painlev\'{e} line element for the Reissner-Nordstr\"{o}m black hole without topological defects \cite{perkih}. Again, these equations are modified when the particle’s self-gravitation is taken into account. It is not difficult to see that the dragging angular velocity of the particle at the horizon  vanishes
\begin{equation}
\Omega'_{+,bp}=\frac{a \tilde{\Delta}(r_{+})}{r^{2}b^{2}p^{2}-a^{2}\tilde{\Delta}(r_{+})}\bigg|_{r=r'_{+}}=0,
\end{equation}
where $\tilde{\Delta}=1-2(M-\omega)/r+Q^{2}/r^{2}$ and $r'_{+}=(M-\omega)+\sqrt{(M-\omega)^{2}-Q^{2}}$. In terms of the action, using Eq. \eqref{g2} with positive sign, it follows that
\begin{equation}
\text{Im}\,S=\text{Im}\,\int_{r_{in}}^{r_{out}}\int_{bp^{2}\,M}^{bp^{2}(M-\omega)}\frac{\sqrt{r^{2}b^{2}p^{2} -a^{2}\tilde{\Delta}}}{bp\left(r-\sqrt{2M'r^{2}-Q^{2}}\right)}\mathrm{d}(bp^{2}M')d r.
\end{equation} 

We can solve this integral by using the Feynman's prescription $i \epsilon$ for positive energy solution, $\omega \to \omega-i\epsilon$, and obtain $M\to M-i\epsilon$, with  $f(M')$ given by
\begin{equation}
f(M')=\frac{\sqrt{r^{2}b^{2}p^{2}-a^{2}\tilde{\Delta}}-bp(r-\sqrt{2M'r^{2}-Q^{2}})}{\sqrt{r^{2}b^{2}p^{2}-a^{2}\tilde{\Delta}}}.
\end{equation}

Using the fact that the radial coordinates ranges from  $r_{in}=M+\sqrt{M^{2}-Q^{2}}$ to $r_{out}=(M-\omega)+\sqrt{(M-\omega)^{2}-Q^{2}}$, solving the integral we find
\begin{eqnarray}\nonumber
\text{Im}\,S
& = &\text{Im}\left[\int_{r_{in}}^{r_{out}}\left(\int_{bp^{2}M}^{bp^{2}(M-\omega)}\,-i\pi \delta \left(1-f(M')\right)\,\mathrm{d}(b\,p^{2}M')\right)\mathrm{d}r\right]\\\nonumber
& = & b\,p^{2}\int_{M+\sqrt{M^{2}-Q^{2}}}^{(M-\omega)+\sqrt{(M-\omega)^{2}-Q^{2}}}\left(-\pi\,r\right)\mathrm{d}r\\\nonumber
& = & \pi b\,p^{2}\left[M^{2}-(M-\omega)^{2}+M\sqrt{M^{2}-Q^{2}}-(M-\omega)\sqrt{(M-\omega)^{2}-Q^{2}}\right].
\end{eqnarray}

We can now use this result and calculate  the tunneling probability of outgoing particles
\begin{eqnarray}
\Gamma=e^{{{-2\pi b\,p^{2}\left(M^{2}-(M-\omega)^{2}+M\sqrt{M^{2}-Q^{2}}-(M-\omega)\sqrt{(M-\omega)^{2}-Q^{2}}\right)}}}.
\label{tunneling rate}
\end{eqnarray}

The outgoing particle starts from just inside the horizon $r_{in}=M+\sqrt{M^{2}-Q^{2}}-\epsilon$ and traverses just outside the horizon $r_{out}=(M-\omega)+\sqrt{(M-\omega)^{2}-Q^{2}}+\epsilon$. We can rewrite the last result in terms of the change of Bekenstein-Hawking entropy before and after the emission
\begin{equation}
\Gamma \sim e^{-2\,\mbox{Im}\,S}=e^{\pi\,b\,p^{2}(r_{out}^{2}-r_{in}^{2})}=e^{\Delta S_{BH,bp}},
\end{equation}
where $\Delta S_{BH,bp}=S_{BH,bp}(M-\omega)-S_{BH,bp}(M)$. Using the Taylor series expansion of $\Delta S_{BH,bp}$ to the first order in $\omega$ and the Boltzmann factor $\Gamma=\exp(-\beta_{H} b\,p^{2}\,\omega)$ the Hawking temperature reads
\begin{equation}
T_{H}=\frac{\sqrt{M^{2}-Q^{2}}}{2\pi\left(M+\sqrt{M^{2}-Q^{2}}\right)^{2}}.
\label{tem1}
\end{equation}

Which is the Hawking temperature for Reissner-Nordstr\"{o}m black hole without a topological defects, in full agreement with Parikh's results \cite{perkih}. Furthermore, the emission spectrum is no longer purely thermal but has some corrections, due to the energy conservation law.

\section{Tunneling of massive particles from Reissner-Nordstr\"{o}m  black holes} 

In the case of tunneling process of massive particles, it is
clear that the trajectory of a massive particle cannot be described by light-like
geodesics. This problem was solved using the de Broglie's
hypothesis \cite{zhang-zhao}. We can consider the outgoing particle as a massive shell whose phase velocity $v_{phase}$ and group velocity $v_{g}$ are given by  
\begin{equation}
v_{phase}=\frac{1}{2}v_{g};\,\,\,\,v_{phase}=\frac{\mathrm{d}r}{\mbox d t},\,\,v_{g}=\frac{\mbox d r_{c}}{\mbox d t}
\label{vg}
\end{equation}
here $r_{c}$ denotes the radial position of the particle described by the weave packet $\Psi(r,t)$. Since the tunneling process is an instantaneous effect, the dragged Reissner-Nordstr\"{o}m cosmic string black hole metric \eqref{r-n}, satisfies Landau's condition of the coordinate clock synchronization with the coordinate time difference of these two events given by
\begin{equation}
\mathrm{d}t=-\frac{g_{01}}{\hat{g}_{00}}\mathrm{d}r_{c},\,\,\,\,\,\,\,\,(\mathrm{d}\theta=0).
\end{equation}

This metric is well behaved at $r_{+}=M+\sqrt{M^{2}-Q^{2}}$ and satisfy Landau's condition of coordinate clock synchronization. The phase velocity is given by
\begin{equation}
v_{g}=\frac{b\,p\,\Xi}{\sqrt{\left(1-\frac{\Xi}{r^{2}}\right)\left(r^{4}b^{2}p^{2}-a^{2}\Xi \right)}},
\end{equation}
where $\Xi=r^{2}-2Mr+Q^{2}$. The group velocity on the other hand is given by
\begin{equation}
\dot{r}=\frac{bp\Xi}{2\sqrt{\left(1-\frac{\Xi}{r^{2}}\right)\left(r^{4}b^{2}p^{2}-a^{2}\Xi \right)}}.
\end{equation}

When a charged particle with energy $\omega$, and charge $q$ is tunneld out, the mass, angular momentum, and the charge of the black hole, will be reduced to $ M-\omega$, $(M-\omega)a$ and $Q-q$, respectively. The corresponding Komar's charge of the black hole also decreases by a factor of $bp^{2}$ due to the presence of topological defects, $Q_{bp}=bp^{2}Q$. And clearly, when a particle tunnels out from the black hole, the charge of the black hole reduces to $bp^{2}(Q-q)$. However, in this scenario we also have to take into account the effect of the electromagnetic field $-(1/4) F_{\mu\nu}F^{\mu\nu}$ described by $A_{\mu}$. Therefore, the Hamilton's equation reads
\begin{eqnarray}
\dot{r}
& = & \frac{\mathrm{d}H}{\mathrm{d}P_{r}}\bigg|_{(r;\varphi,P_{\varphi},A_{t}, P_{A_{t}})}=bp^{2}\frac{\mathrm{d}(M-\omega)}{\mathrm{d}P_{r}},\\
\dot{\varphi}
& = & \frac{\mathrm{d}H}{\mathrm{d}P_{\varphi}}\bigg|_{(\varphi;r,P_{r},A_{t}, P_{A_{t}})}=abp^{2}\Omega'_{+,bp}\frac{\mathrm{d}(M-\omega)}{\mathrm{d}P_{\varphi}},\\
\dot{A_{t}}
& = & \frac{\mathrm{d}H}{\mathrm{d}P_{A_{t}}}\bigg|_{(A_{t};r,P_{r},\varphi,P_{\varphi})}=bp^{2}\Phi_{H}\frac{\mathrm{d}(Q-q)}{\mathrm{d}P_{A_{t}}}
\end{eqnarray}

When the particle's self-gravitation is taken into account, and make use of the fact that the dragging angular velocity $\Omega'_{+,bp}(r=r'_{+})$, vanishes at the horizon $r'_{+}=(M-\omega)+\sqrt{(M-\omega)^{2}-(Q-q)^{2}}$, therefore the electric potential at the event horizon in the dragging coordinate system will be
\begin{equation}
\Phi'_{H}=\frac{(Q-q)}{(M-\omega)+\sqrt{(M-\omega)^{2}-(Q-q)^{2}}}.
\label{electricpotential}
\end{equation}

Since the metric (\ref{r-n}) is independent of the coordinate $\varphi $, and $ \dot{A}_{t}$, we can complitly  eliminate these two degrees of freedom by writing the
action of the charged massive particle in the form
\begin{eqnarray}
\mbox{Im}\,S
& = & \mbox{Im}\int\limits_{r_{in}}^{r_{out}}\,\left(P_{r}\dot{r}-P_{\varphi}\dot{\varphi}-P_{A_{t}}\dot{A_{t}}\right)\mathrm{d}t\\\nonumber
& = & \mbox{Im}\int\limits_{r_{in}}^{r_{out}}\left[\int\limits_{(0,0,0)}^{(P_{r},P_{\varphi},P_{A_{t}})}\left(\dot{r}\mathrm{d}P'_{r}-\dot{\varphi}\mathrm{d}P'_{\varphi}-\dot{A}_{t}\mathrm{d}P'_{A_{t}}\right)\right]\frac{\mathrm{d}r}{\dot{r}}.
\label{action}
\end{eqnarray}

Using the Feynman's prescription $i \epsilon$ for positive energy solution $\omega \to \omega-i\epsilon$, and charge $q \to q-i\epsilon$ we obtain $M'\to M'-i\epsilon$ and $Q'\to Q'-i\epsilon$, here we regard $\omega$ and $q$ as  constants and $M'$ and $Q'$ as variables. The outgoing particle starts from just inside the horizon $r_{in}=M+\sqrt{M^{2}-Q^{2}}-\epsilon$ and traverses just outside the horizon $r_{out}=(M-\omega)+\sqrt{(M-\omega)^{2}-(Q-q)^{2}}+\epsilon$. For the imagionary part of the action we find
\begin{eqnarray}\nonumber
\mbox{Im}\,S 
& = & \mbox{Im} \int\limits_{r_{in}}^{r_{out}}\int\limits_{bp^{2}(M,Q)}^{bp^{2}(M-\omega,Q-q)}\left[\mathrm{d}\left(bp^{2}M'\right)-\Phi'_{H}\, \mathrm{d}\left(bp^{2}Q'\right)\right]\frac{\mathrm{d}r}{\dot{r}}\\ \nonumber
& = & \mbox{Im} \int\limits_{r_{in}}^{r_{out}}\int\limits_{bp^{2}(M,Q)}^{bp^{2}(M-\omega,Q-q)}\frac{2\sqrt{(1-\frac{\tilde{\Xi}}{r^{2}})(r^{4}b^{2}p^{2}-a^{2}\tilde{\Xi})}}{bp\tilde{\Xi}}\left[\mathrm{d}(bp^{2}M')-\frac{Q'\,\mathrm{d}(bp^{2}Q')}{M'+\sqrt{M'^{2}-Q'^{2}}}\right]\mathrm{d}r \\
& = & \frac{\pi\,b\,p^{2}}{2} \left(r_{in}^{2}-r_{out}^{2}\right)
\end{eqnarray}
where $\tilde{\Xi}=r^{2}-2(M-\omega)r+(Q-q)^{2}$. The radial coordinates ranges from $r_{in}=M+\sqrt{M^{2}-Q^{2}}$ to $r_{out}=(M-\omega)+\sqrt{(M-\omega)^{2}-(Q-q)^{2}}$, for the tunneling rate it follows that
\begin{eqnarray}
\Gamma \sim e^{\pi\, b\,p^{2}\left[\left(M-\omega+\sqrt{(M-\omega)^{2}-(Q-q)^{2}}\right)^{2}-\left(M+\sqrt{M^{2}-Q^{2}}\right)^{2}\right]}
\label{tunneling rate}
\end{eqnarray}

Using the determinant of the metric (\ref{Reissner}) and the corresponding surface area of the Reissner-Nordstr\"{o}m black hole, before and after the emission of the particle, one can check that the tunneling rate in terms of the change of Bekenstein-Hawking entropy $\Delta S_{BH,bp}=S_{BH,bp}(M-\omega, Q-q)-S_{BH,bp}(M,Q)$, reads
\begin{equation}
\Gamma \sim e^{-2\mbox{Im}\,S}=e^{\pi\,b\,p^{2}(r_{out}^{2}-r_{in}^{2})}=e^{\Delta S_{BH,bp}}.
\end{equation}

Setting $q=0$, and doing a Taylor series expansion of (\ref{tunneling rate}) to the first order in $\omega$ and using the Boltzmann factor $\Gamma=\exp{\left(-\beta_{H}bp^{2}\omega \right)}$, we recover the correct Hawking temperature \eqref{tem1} for massive particles.

\section{Conclusion}
In this paper, we have investigated the Hawking radiation of massless particles using a semi-classical WKB approximation for the case of Schwarzschild and Reissner-Nordstr\"{o}m black holes with topological defects. By introducing a Painlev\'{e} type coordinates and taking into account that the ADM mass, energy, and angular momentum, shifts due to the presence of a cosmic string and a global monopole we have calculated the tunneling probability of outgoing particles. By comparing the tunneling rate with the Boltzmann factor we have calculated the Hawking temperature for these black holes. As a result, we have shown that the Hawking radiation remains uneffected by the presence of topological defects in both cases. For the case of charged massive particles, we have also recovered the same Hawking temperature for Reissner-Nordstr\"{o}m black hole, here the radial trajectory can be determined by using the de Broglie's hypothesis. Finally, the results in this paper extend the Perkih's work and are consistent with an underlying unitary theory.

\vspace*{0.3cm}
\section*{Acknowledgement}
The author would like to thank the editor and the anonymous reviewers for the very useful comments and suggestions which help us improve the quality of this paper.

\end{document}